\documentclass[prb,showpacs,superscriptaddress,twocolumn]{revtex4}%
\usepackage{amsfonts}
\usepackage{amsmath}
\usepackage{amssymb}
\usepackage{graphicx}%
\begin{document}
\title{First-principles calculations of Cu(001) thin films: quantum size effect in
surface energetics and surface chemical reactivities}
\author{Bo Sun}
\affiliation{Institute of Applied Physics and Computational
Mathematics, P.O. Box 8009, Beijing 100088, P.R. China}
\author{Ping Zhang}
\thanks{Electronic address: zhang\_ping@iapcm.ac.cn}
\affiliation{Institute of Applied Physics and Computational
Mathematics, P.O. Box 8009, Beijing 100088, P.R. China}
\author{Suqing Duan}
\affiliation{Institute of Applied Physics and Computational
Mathematics, P.O. Box 8009, Beijing 100088, P.R. China}
\author{Xian-Geng Zhao}
\affiliation{Institute of Applied Physics and Computational
Mathematics, P.O. Box 8009, Beijing 100088, P.R. China}
\author{Junren Shi}
\affiliation{Institute of Physics, Chinese Academy of Sciences, Beijing 100080, P.R. China}
\author{Qi-Kun Xue}
\affiliation{Institute of Physics, Chinese Academy of Sciences,
Beijing 100080, P.R. China} 

\begin{abstract}
First-principles calculations of Cu(001) free-standing thin films have been
performed to investigate the oscillatory quantum size effects exhibited in
surface energy, work function, atomic relaxation, and adsorption energy of the
cesium adsorbate. The quantum well states have been shown and clarified at
particular $k$-points corresponding to the stationary extrema in bulk
Brillouin zone, which are in good agreement with experimental observations.
The calculated surface energetics and geometry relaxations are clearly
featured by quantum oscillations as a function of the film thickness of the
film with oscillation periods characterized by a superposition of long and
short length scales. Furthermore, we have investigated Cs adsorption onto
Cu(001) thin films as a function of the film thickness. Our systematic
calculated results clearly show the large-amplitude quantum oscillations in
adsorption energetics, which may be used to tailor catalysis, chemical
reactions and other surface processes in nanostructured materials.

\end{abstract}
\maketitle

\section{Introduction}

When the thickness of thin metal films approaches the nanoscale, the
oscillatory quantum size effects (QSEs) associated with electronic confinement
and interference will occur\cite{Pag,Tes,Mey,Kaw} due to the splitting of the
energy-level spectrum into subbands normal to the plane of the films, i.e.,
the quantum well (QW) states. These QW states lead to strongly modified
physical properties and thus have been the subject of numerous experimental
investigations in recent years\cite{Chiang2000,Milun}. For example, the QW
states are found to be responsible for an unusual metallic film growth
pattern\cite{Hupalo,Su2001,Otero2002,Hong2003}, and for the
thickness-dependent stability\cite{Luh2001} observed in the experiment. The QW
states are directly connected to the oscillation in the exchange coupling
between two magnetic materials across a nonmagnetic spacer layer of various
thickness\cite{Qiu2002}, and to giant
magnetoresistance\cite{Baibich1988,Ortega1992,Bruno1993,Edwards1991}.
Moreover, the QW states also give rise to an oscillatory phonon-electron
coupling as the film thickness varies, and thus affect the transition
temperature of the superconductivity\cite{Xue1,Xue2}.

Experimentally, the characterization of the QW states are commonly measured
using angle-resolved photoemission spectroscopy (ARPES) and the scanning
tunneling microscopy (STM). ARPES can be used to study the band structure
along any direction of the surface Brillouin zone (BZ), while STM offers the
possibility to study local structures, such as islands, chains, dots, etc.
Using the scanning tunneling spectroscopy (STS) technique, in which the
differential conductance ($dI/dV$) is measured, one can reliably determine the
energy of quantized electronic states in the range of approximately 1 eV below
and above the Fermi level. Theoretically, a number of approaches have been
used in the past in order to describe the electronic properties, in particular
the QW states, in ultrathin metallic films. Quasi-one-dimensional models, such
as a square well potential\cite{Kralj} or the phase accumulation model
(PAM)\cite{Lindgren1987}, have been successfully used to interpret the
energies of QW states. More sophisticated methods have also been used, such as
the tight-binding approach\cite{Smith1994} and layer-Korringa-Kohn-Rostoker
approach\cite{Moruzzi1978,Wang2006}. In a few systems the QW states have been
investigated by self-consistent density-functional
calculations\cite{Fei,Kie,Cir,Boe,Carlsson2000,Wei2002,Wei2003,Lazic2005}.
There are strong reasons to use the \textit{ab initio} methods. First, there
are no adjustable parameters, and a wide range of calculated structural and
electronic properties offer the possibility of a detailed comparison with
experiments. Also, quantities such as the expected STM profiles and the
amplitudes of the wave functions of the QW states, which cannot be obtained in
simple approaches, can be calculated.

In this paper we report our first-principles calculations of the QSE in a
specific QW system, i.e., the Cu(001) freestanding thin films. Previous QSE
studies concerning Cu(001) are mainly focused on the oscillations in magnetic
interaction between the two ferromagnetic layers in fcc $M$/Cu/fcc $M$ (001)
sandwich structures where $M$ denotes the ferromagnetic material. It has been
demonstrated that the Cu(001) films have a long length scale of 5.6 monolayer
(ML) and a short length scale of 2.6 ML of oscillation periods for magnetic
coupling in the [001] direction, corresponding to spanning vectors at the
belly and neck of the bulk Cu Fermi surface
respectively\cite{D-J,Cur,Klsges,Kawakami}. Recent experimental efforts have
been focused on the new band structure properties of QW states in Cu(001)
system\cite{Wu01,Danese,J-M,Wu02,Rotenberg,Wu03,Ling,Chiara,Bisio}. On the
other side, the QSE of Cu(001) associated with its energetics was scarcely
considered up to now. In particular, there is no clear experimental or
theoretic evidence of the interplay between the different oscillation periods
of Cu(001) films by the belly and neck extrema in bulk Cu Fermi surface. In
this paper we present a detailed first-principles study of the surface
energetics of the Cu(001) free standing thin films. The QW states
corresponding to the stationary extrema in bulk BZ are studied in detail. The
oscillations of the energetics versus the Cu(001) film thickness are
identified and the corresponding oscillation periods are explained. We find
that the quantum interference of the QW states with different in-plane wave
vectors result in a superposion between long- and short-length oscillating periods.

The other purpose in this paper is to investigate the QSE character in surface
adsorption energetics for a representative system in order to shed light on
the effect of the QW states on the surface reactivities. Since the adsorption
property is closely characterized by the chemical bonding between the
adsorbate and the surface of the substrate, thus when the substrate is
ultrathin, the QSE in the substrate will also influence the behavior of the
surface adsorption. Here we address a particular adsorbate system, i.e.,
Cs/Cu(001), as a case study to manifest the QSE in surface adsorption
properties. Due to its simple electronic structure and active chemical
properties which is intrinsic for alkali metals, Cs is a unique adsorbate on
metal surfaces, and has been extensively studied. Concerning Cs adsorption on
Cu metal surfaces, the recently published
articles\cite{Bauer,Ogawa,Borisov,Gauyacq,Corriol,Hofe} mainly focus on the
investigation of electronic, dynamic, and geometry properties of Cs layers on
Cu(111). The nature of interaction between the Cs adatom and Cu(001) film has
been studied both via first-principles pseudopotential
calculations\cite{Chudinov} or through and a jellium model
approach\cite{Wu1990}.

Compared to the previous Cs/Cu(001) work, as mentioned above, our present
emphasis is on QSE in surface adsorption energetics, instead of giving a
general overview of the nature of alkali-metal-atom adsorption on metal
surfaces. In particular, although it has been well known that due to charge
transfer from the adsorbed alkali atom to the substrate the alkali adatoms
become partially charged\cite{Wandelt}, the dependence of this charge transfer
process on the thickness of the ultrathin substrate film is yet to be
understood with high interest. We notice there are emerging some high-quality
experimental data of QSE in charge transfer, via observation of the absorbed
work function and photoemission as a function of the thickness of the
ultrathin substrate film\cite{Qiu2007}. Further measurements related with the
QSE effect of the adsorption are expected to be systematically reported
afterwards due to the availability of the high-quality quantum metal thin
films. From this aspect a thorough theoretical investigation is necessary and
will be helpful for experimental references in near future. Our results show
that in the ultrathin Cu(001) films, the surface adsorption properties display
well-defined QSEs.

This paper is organized as follows: In Sec. II, the \textit{ab initio} based
method and computational details are outlined. In Sec. III, the band structure
and the properties of QW states at the belly and neck points in bulk BZ are
presented. In Sec. IV, the surface properties of the Cu(001) films, surface
energy, work function, and interlayer relaxation, as a function of the
thickness of the films, are presented and discussed. In Sec. V, the adsorption
properties of alkali-metal Cs on Cu(001) surface is discussed in detail by
presenting the sensitivity of the adsorption energy to the thickness of the
Cu(001) films. Finally, Sec. IV contains a summary of the work and our conclusion.

\section{Computational method}

The calculations were carried out using the Vienna \textit{ab initio}
simulation package\cite{Vasp} based on density-functional theory with PAW
pseudopotentials\cite{Paw} and plane waves. In the present film calculations,
free-standing Cu(001) films by the so-called \textit{repeated slab} geometries
were employed. This scheme consists in in the construction of a unit cell of
an arbitrarily fixed number of atomic layers identical to that of the bulk in
the plane of the surface (defining the two dimensional cell), but
symmetrically terminated by an arbitrarily fixed number of empty layers (the
\textquotedblleft\textit{vacuum}\textquotedblright) along the direction
perpendicular to the surface. In the present study we have fixed the whole
vacuum region equal to 20 \AA , which is found to be sufficiently convergent.
The two dimensional unit cell of the fcc Cu(100) surface is a square of edge
$a/\sqrt{2}$ with basis vectors $a_{1}=a/2(1,\bar{1},0)$ and $a_{2}%
=a/2(1,1,0)$ where $a$ is the Cu bulk lattice constant. The corresponding
surface BZ is a square with two high-symmetry directions $\bar{\Gamma}$%
-$\bar{M}$ and $\bar{\Gamma}$-$\bar{X}$. During our slab calculations the BZ
integration was performed using the Monkhorse-Pack scheme\cite{Pack} with a
$11\times11\times1$ $k$-point grid, and the plane-wave energy cutoff was set
$270$ eV. Furthermore, the generalized gradient approximation (GGA) with PW-91
exchange-correlation potential has been employed with all atomic
configurations fully relaxed. First the total energy of the bulk fcc Cu was
calculated to obtain the bulk lattice constants. The calculated lattice
constant is $a=3.639$ \AA , comparable with experimental\cite{Kittel} values
of $3.61$ \AA , respectively. The use of larger $k$-point meshes did not alter
these values significantly. A Fermi broadening of 0.1 eV was chosen to smear
the occupation of the bands around $E_{F}$ by a finite-$T$ Fermi function and
extrapolating to $T=0$ K.

\section{Band structure and quantum well states}%

%TCIMACRO{\TeXButton{TeX field}{\begin{figure}[tbp]
%\begin{center}
%\includegraphics[width=1.0\linewidth]{fig1.eps}
%\end{center}
%\caption
%{(Color online) (a) GGA energy bands and density of electron states (right panel) of fcc bulk Cu; (b) The Cu Fermi surface showing
%the belly and neck regions. The bulk and surface BZ are also depicted; (c) Calculated (GGA) energies at $\overline
%{\mathrm{\Gamma}}%
%$ in Cu(001) thin films as a function of thickness, with the energy set to zero at the Fermi level. The right
%panel plots the bulk energy dispersion in the [001] direction; (d) Calculated (GGA) energies at $\overline
%{\mathrm{P}}%
%$ in Cu(001) thin films as a function of thickness, with the energy set to zero at the Fermi level. The right
%panel plots the bulk energy dispersion in the [001] direction. The dotted line denotes Fermi level.}
%\label{fig1}
%\end{figure}}}%
%BeginExpansion
\begin{figure}[tbp]
\begin{center}
\includegraphics[width=1.0\linewidth]{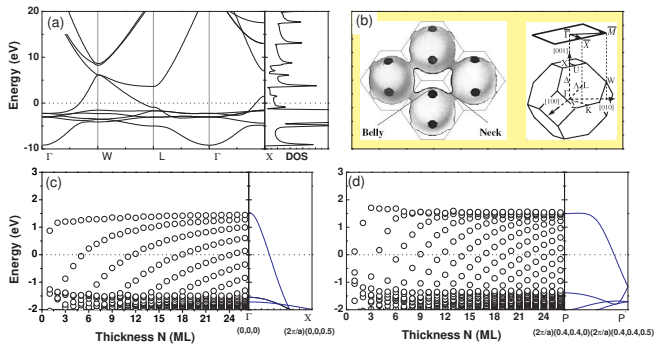}
\end{center}
\caption
{(Color online) (a) GGA energy bands and density of electron states (right panel) of fcc bulk Cu; (b) The Cu Fermi surface showing
the belly and neck regions. The bulk and surface BZ are also depicted; (c) Calculated (GGA) energies at $\overline
{\mathrm{\Gamma}}%
$ in Cu(001) thin films as a function of thickness, with the energy set to zero at the Fermi level. The right
panel plots the bulk energy dispersion in the [001] direction; (d) Calculated (GGA) energies at $\overline
{\mathrm{P}}%
$ in Cu(001) thin films as a function of thickness, with the energy set to zero at the Fermi level. The right
panel plots the bulk energy dispersion in the [001] direction. The dotted line denotes Fermi level.}
\label{fig1}
\end{figure}%
%EndExpansion
We first studied the properties of electronic structures for bulk Cu and
Cu(001) films. For this we have used two kinds unit cells of crystal lattice.
Namely, unit cell I is defined via introducing the usual fcc basis vectors
$\mathbf{a}_{1}=(a/2)(\mathbf{j}+\mathbf{k})$, $\mathbf{a}_{2}%
=(a/2)(\mathbf{i}+\mathbf{k})$, and $\mathbf{a}_{3}=(a/2)(\mathbf{i}%
+\mathbf{j})$, while unit cell II is defined by choosing Cu(001) as the basal
plane, i.e., $\mathbf{a}_{1}=(a/2)(\mathbf{i}-\mathbf{j})$, $\mathbf{a}%
_{2}=(a/2)(\mathbf{i}+\mathbf{j})$, and $\mathbf{a}_{3}=a\mathbf{k}$. The
volume of second unit cell is twice as that of the first one and it's
convenient to extend to slab calculation since $\mathbf{a}_{3}$ is normal to
Cu(001) surface. Correspondingly, the reciprocal lattice basis vectors are
$\mathbf{b}_{1}=(2\pi/a)(-\mathbf{i}+\mathbf{j+k})$, $\mathbf{b}_{2}%
=(2\pi/a)(\mathbf{i-j}+\mathbf{k})$, and $\mathbf{b}_{3}=(2\pi/a)(\mathbf{i+j}%
-\mathbf{k})$ for unit cell I and $\mathbf{b}_{1}=(2\pi/a)(\mathbf{i}%
-\mathbf{j})$, $\mathbf{b}_{2}=(2\pi/a)(\mathbf{i}+\mathbf{j})$, and
$\mathbf{b}_{3}=(2\pi/a)\mathbf{k}$ for unit cell II. Figure 1(a) shows the
band structure and the DOS of bulk fcc Cu (unit cell I). The highly dispersive
$s$-$p$ band, typical of noble metals, is present. The more intense $d$ band
region is located between 2.0 and 4.5 eV below Fermi surface and can be
measured by a large rise of the intensity in photoemission
experiments\cite{Qiu2002}. The $s$-$d$ hybrid band region with two less
intense dispersive states can also be seen in Fig. 1(a). The Fermi surface of
bulk Cu and the Brillouin zone for unit cell I are schematically shown in Fig.
1(b). Of particular interest for Cu(001) thin film studies are the two
extremal points, i.e., the belly and neck points at the bulk Fermi surface.
Previous extensive studies have shown that the stationary spanning vectors
connecting the belly and neck points respectively play a key role in
determining the oscillatory behavior of interlayer magnetic exchange
interaction $J_{M/M}$ in fcc $M$/Cu/$M$ (001) sandwich structures where $M$
denotes the ferromagnetic material. The belly extrema was found to correspond
to long oscillating period in $J_{M/M}$ while the neck extrama corresponds to
short period.

One inconvenience in discussing properties of Cu(001) film via the bulk Fermi
surface calculated from unit cell I lies in the fact that none of three basis
vectors of unit cell I is normal to (001) surface, which leads to an oblique
projection of bulk BZ onto (001) surface BZ. On the contrary, by use of unit
cell II the discussions, particularly for QW states of (001) thin films,
become very convenient. Beginning with this unit cell construction, the belly
point in bulk BZ is still projected to $\overline{\mathrm{\Gamma}}$ point,
like with unit cell I. Whereas the neck point, after projection, turns to be
at $\overline{\mathrm{\Gamma}}\overline{\mathrm{X}}$ line. In this paper the
neck point will be named $\mathrm{P}$ and $\overline{\mathrm{P}}$ for bulk and
surface BZ (with respect to unit cell II) respectively. Our band structure
calculation gives the position [in coordinates of ($k_{x},k_{y},k_{z}$)] of
$\mathrm{P}$ to be $(2\pi/a)(0.4,0.4,0.0)$.

Now we turn to study the electronic structure of Cu(001) thin film with focus
on QW states. Since the previous studies\cite{Qiu2002} have shown that the
long and short oscillating periods in interlayer magnetic exchange interaction
originate from band dispersions at $\overline{\mathrm{\Gamma}}$ and
$\overline{\mathrm{P}}$ points respectively, we expect the other intrinsic
film properties such as film energetics, atomic structure relaxations, and
even surface chemistry are closely related with the QW states at these two
kinds of stationary $k$-points. Figure 1(c) shows the energies of the QW
states at $\overline{\Gamma}$ point as a function of the film thickness
without interlayer relaxation. The energy zero is set at the Fermi level of
each film. Interlayer relaxation effect is also studied and it is found that
the overall thickness dependence of the energies are similar to that without
relaxation. For comparison, also plotted in Fig. 1(c) (right panel) is the
energy dispersion in the bulk along [001] ($\Gamma\rightarrow X$) direction,
which corresponds to the center $\overline{\Gamma}$ of (001) surface BZ and
determines the energy range for the QW states shown in the left panel. One can
see from Fig. 1(c) that as the thickness of the film increases, the energy of
a given QW state also increases. When the thickness of the film is increased
to be about 5.0 monolayers, then a quantum well state, with the energy
crossing the Fermi level, occurs. The next energy crossing with the Fermi
level occurs at the film thickness of $\sim$11 monolayers. Our calculated
result of quantum well states is in good agreement with the recent
experimental ARPES measurement\cite{Qiu2002}, in which the photoemission
intensity as a function of energy for fixed Cu thickness and as a function of
Cu thickness for fixed energy clearly show the existence of the QW states of
the $sp$ electrons in the Cu layer. Fig. 1(d) shows (left panel) the energies
of the QW states at $\overline{\mathrm{P}}$ point as a function of film
thickness without interlayer relaxation. The energy zero is set at the Fermi
level of each film. Again the bulk energy dispersion along [001] direction
(starting from $\mathrm{P}$ point). One can see that for every incremental
increase in the film thickness of about 2.6 ML, then a new quantum well state
crosses the Fermi level. A comparison between the left panel and right panel
in Fig. 1(c) [or (d)] shows that the bulk electron band of Cu works very well
for QW states as the Cu thickness is greater than 5 ML, which agrees well with
the experimental results\cite{Kawakmi1998}.

The quantitative understanding of the QW states showed in Figs. 1(c)-(d) is
usually analyzed using the so-called phase accumulation model
(PAM)\cite{Ech,Smith}. Here the free standing Cu(001) film is considered as a
quantum well confining electrons between the two vacuums in the slab. Since
the system is invariant on translation parallel to the (001) plane, the
in-plane wavevector $\mathbf{k}_{\parallel}\equiv(k_{x},k_{y})$ is a good
quantum number. Thus, for a given $\mathbf{k}_{\parallel}$, the quantization
condition for an electron state in such a well is given by
\begin{equation}
2k_{n}^{\perp}Nd+2\phi=2\pi n,\tag{1}%
\end{equation}
where $N$ is the number of atomic layers in the film, $d=a/2$ the interlayer
spacing, $\phi_{B}$ is the phase gain of the electron wave function upon
reflection at the the film-vacuum interface, $n$ is the number of
half-wavelengths confined inside the QW and $k_{n}^{\perp}$ describes the Cu
electron wavevector component along [001] direction for the $n$th QW state.
Equation (1) has been successfully used to model QW states in metal thin
films, and its validity has been rigorously tested by full-scale
first-principles calculations for some special systems\cite{Wei2002}. It
should be addressed that QW states will not only form at the $\overline
{\mathrm{\Gamma}}$ point but also form in a large part of the surface BZ, and
the quantization condition (1) may be applied throughout the entire zone.
Using Eq. (1) one can calculate the periodicity for the QW states crossing the
Fermi level, $\Delta N=\pi/[k_{F}^{\perp}(\mathbf{k}_{\parallel})d]$, where
$k_{F}^{\perp}(\mathbf{k}_{\parallel})$ is the bulk Fermi wavevector along
[001] direction for a given in-plane wavevector $\mathbf{k}_{\parallel}$. For
the $\overline{\mathrm{\Gamma}}$ point ($\mathbf{k}_{\parallel}=0$), from the
right panel in Fig. 1(c), one can see that the upper branch of the bulk $sp$
band runs through $\sim33\%$ of the BZ, $k_{\bot}^{f}=0.33\pi/c$. One gets
$\Delta N=6$. Therefore a new QW state of $\mathbf{k}_{\parallel}=(0,0)$
occurs every $5.6$ ML increase of the film thickness, which is verified in the
left panel in Fig. 1(c) that an energy branch moves down, crossing the Fermi
level for every incremental increase in the film thickness of $6$ layers.
Similarly for $\overline{\mathrm{P}}$ point corresponding to $\mathbf{k}%
_{\parallel}=(2\pi/a)(0.4,0.4)$ in bulk BZ, one can derive from the right
panel in Fig. 1(c) $k_{\bot}^{f}=0.75\pi/c$. In this case, Eq. (1) gives the
periodicity for the QW states to be $\Delta N=2.6$. Therefore, Figs. 1(c)-(d)
reveal that the periodic behavior of the Cu(001) QW states is essentially
governed by the bulk Cu electronic structure properties. In particular, the
extremal points of the Fermi surface and the extremal points of the QW
branches close to the Fermi level occur at the same $\mathbf{k}_{\parallel}$,
i.e., at $\mathbf{k}_{\parallel}=(0,0)$ in the bulk BZ corresponding to
$\overline{\Gamma}$ in the (001) surface BZ and at the neck extrema
$\mathbf{k}_{\parallel}=(2\pi/a)(0.4,0.4)$ both in the bulk BZ and in the
surface BZ. This relationship between the Fermi surface and the quantum-well
dispersion turns out to be very fruitful for the understanding of the
connection between the periods in energetics and surface reactivity discussed below.%

%TCIMACRO{\TeXButton{TeX field}{\begin{figure}[tbp]
%\begin{center}
%\includegraphics[width=1.0\linewidth]{fig2.eps}
%\end{center}
%\caption
%{ The phase shift of the electronic wave function at the film-vacuum interface as a function of energy of QW states at
%(a) $\overline{\mathrm{\Gamma}}$ and (b) $\overline{\mathrm{P}}$ points.}
%\label{fig1}
%\end{figure}}}%
%BeginExpansion
\begin{figure}[tbp]
\begin{center}
\includegraphics[width=1.0\linewidth]{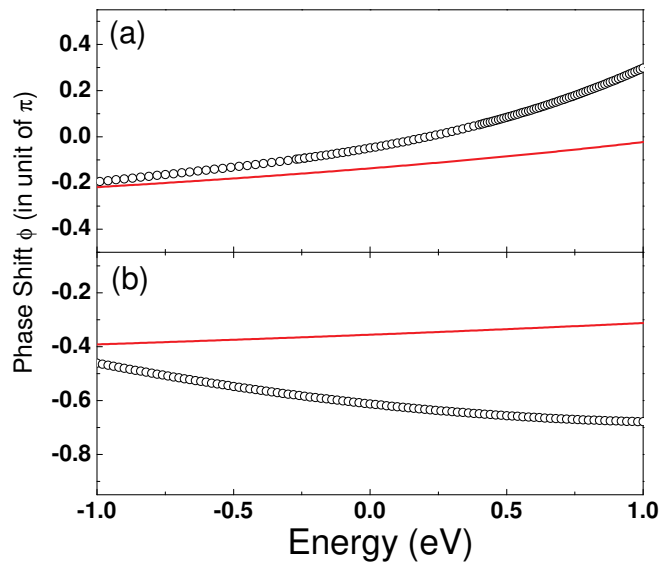}
\end{center}
\caption {(Color online) The phase shift of the electronic wave
function at the film-vacuum interface as a function of energy of QW
states at (a) $\overline{\mathrm{\Gamma}}$ and (b)
$\overline{\mathrm{P}}$ points.} \label{fig1}
\end{figure}%
%EndExpansion
Using the energy data for QW states in Figs. 1(c)-(d), and with the help of
Eq. (1) and the bulk band structure shown in the right panels in Figs.
1(c)-(d), one can deduce the phase shifts $\phi$ for the QW states at
$\overline{\mathrm{\Gamma}}$ and $\overline{\mathrm{P}}$ points. They are
plotted as filled circles in Fig. 2(a)-(b) respectively. Since the energy
range of interest is quite far from the calculated vacuum level of 4.5 eV
(relative to the Fermi level), this phase shift curve is nearly featureless
for most of the energy range, except for the feature near the top of the band.
In general the phase shift depends on the incident (with respect to the
film/vauum interface) energy and momentum of the electron. A simple expression
obtained from the WKB approximation is often used in the literature and has
the form\cite{McRae}
\begin{equation}
\phi(E)=\pi\sqrt{\frac{\text{3.4 (eV)}}{\text{4.5 (eV)}-(E-\hbar
^{2}k_{\parallel}^{2}/2m)}}-\pi.\tag{2}%
\end{equation}
Here 4.5 eV is the Cu work function (in the thin thickness limit), $E$ is the
electron energy measured from the Fermi level and $k_{\parallel}$ is the the
in-plane wavevector of QW. The image potential due to the electron--hole
attraction is like a one-dimensional hydrogen atom. That is where the 3.4 eV
is from which actually equals one quarter of the hydrogen ionization energy.
For the belly of the Cu Fermi surface $k_{\parallel}=0$ while for neck point
we obtain $k_{\parallel}=0.98$ \r{A}$^{-1}$. Equation (2) is plotted as a
dashed curve in Fig. 2. Note that the data points within a given energy range
can come from films of very different thickness. The datasets at both
$\overline{\mathrm{\Gamma}}$ and $\overline{\mathrm{P}}$ vary smoothly as a
function of energy with little scattering, indicating that Eq. (1) does
determine the energies of the QW states quantitatively, provided the phase
shifts are known.%

%TCIMACRO{\TeXButton{TeX field}{\begin{figure}[tbp]
%\begin{center}
%\includegraphics[width=1.0\linewidth]{fig3.eps}
%\end{center}
%\caption
%{The inverse of the energy gap between the highest occupied QW state and the
%lowest unoccupied QW state as a function of Cu(001) film
%thickness. The squares in the figure are the results for $\overline{\Gamma}$
%point in surface BZ, while the circles are for $\overline{\mathrm{P}%
%}$ point (projection of
%bulk Fermi surface's neck onto the (001) plane).}
%\label{fig1}
%\end{figure}}}%
%BeginExpansion
\begin{figure}[tbp]
\begin{center}
\includegraphics[width=1.0\linewidth]{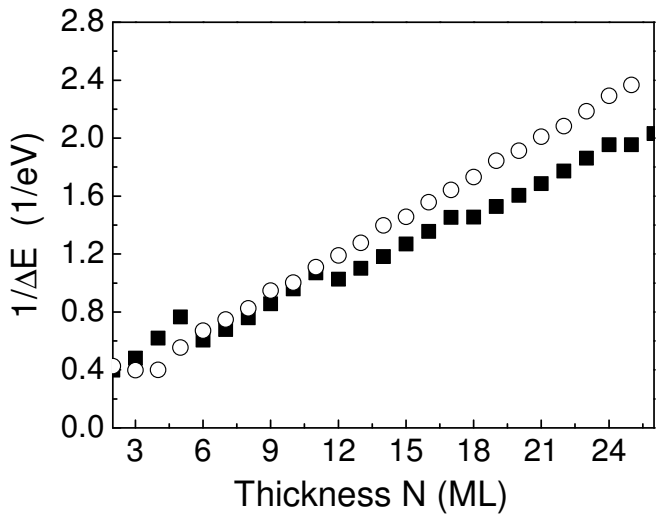}
\end{center}
\caption
{The inverse of the energy gap between the highest occupied QW state and the
lowest unoccupied QW state as a function of Cu(001) film
thickness. The squares in the figure are the results for $\overline{\Gamma}$
point in surface BZ, while the circles are for $\overline{\mathrm{P}%
}$ point (projection of
bulk Fermi surface's neck onto the (001) plane).}
\label{fig1}
\end{figure}%
%EndExpansion
The energies of QW states near the Fermi level can be measured by STS
experiments, in which the local density of states (LDOS) is probed through the
\textrm{d}$I$/\textrm{d}$V$ curve. The distinctive sharp peaks in
\textrm{d}$I$/\textrm{d}$V$ curve are characteristic of the QW states at
different quantum numbers. The most easy-to-see quantity in STS measurement is
the energy gap $\Delta E$ between the highest occupied QW state and lowest
unoccupied QW state. To see the thickness dependence of $\Delta E$, taking the
derivative of Eq. (1) with respect to energy and evaluating it at the Fermi
level for a given $N$, one has\cite{Wei2002}%
\begin{equation}
\frac{1}{\Delta E}\approx\frac{2d}{hv_{F}}N+\frac{1}{2\pi}\phi^{\prime}%
(E_{F})\tag{3}\label{E3}%
\end{equation}
where $v_{F}$ is the Fermi velocity obtained from the slope of the bulk band
at the Fermi level and $\phi^{\prime}(E_{F})$ the energy derivative of the
interface electronic phase shift at the Fermi level. Therefore, the measured
$1/\Delta E$ curve is a linear function of $N$, with a slope connected to
$v_{F}$. Figure 3 shows the calculated $1/\Delta E$ at $\overline
{\mathrm{\Gamma}}$ and $\overline{\mathrm{P}}$ points using the QW energies in
Figs. 1(c)-(d). One can see that the two curves follow a straight line with
different slopes. Also it can be seen that whence a new branch of QW states
crosses the Fermi level, then a kink occurs. Due to the different periods at
$\overline{\mathrm{\Gamma}}$ and $\overline{\mathrm{P}}$, the kinks in these
two curves are also located at different values of $N$. Note that the
intersection of the linear curves with the horizontal axis is not necessarily
at $N=0$, due to the nonzero energy derivative of the interfacial phase shifts
at the Fermi level\cite{Wei2003}.%

%TCIMACRO{\TeXButton{TeX field}{\begin{figure}[tbp]
%\begin{center}
%\includegraphics[width=1.0\linewidth]{fig4.eps}
%\end{center}
%\caption{Calculated electronic density of states at the Fermi energy
%as a function of Cu(001) thin film thickness.} \label{fig5}
%\end{figure}}}%
%BeginExpansion
\begin{figure}[tbp]
\begin{center}
\includegraphics[width=1.0\linewidth]{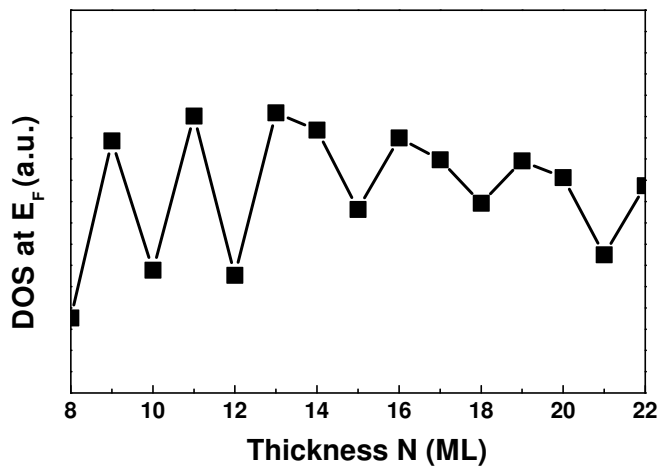}
\end{center}
\caption{Calculated electronic density of states at the Fermi energy
as a function of Cu(001) thin film thickness.} \label{fig5}
\end{figure}%
%EndExpansion
Another central quantity closely related with above discussed QW electronic
structure of Cu(001) thin film is the film electronic DOS at the Fermi level,
$D(E_{F})$. As shown in Fig. 4, the film $D(E_{F})$ with respect to the film
thickness displays well-defined oscillations with oscillation periods
characterized by a superposition of long and short length. Note that in
obtaining Fig. 4 we have increased the $k$-grid for integration to
$35\times35\times1$ for each value of $N$, which ensures the precision of the
result. The oscillations in $D(E_{F})$ give an periodic band energy
contribution to the total energy, thus producing oscillatory surface
energetics and reactivity which will be shown below.

\section{Film energetics and interlayer relaxation}

Figure 3(a) shows the total energy per monolayer $E(N)/N$ as a function of the
number $N$ of layers in the slab. The atoms in the slabs have been fully
relaxed during the calculations. One can see from Fig. 3(a) that with
increasing the thickness, $E(N)/N$ gradually approaches a constant value which
in the limit is equal to the energy per atom in bulk Cu.%

%TCIMACRO{\TeXButton{TeX field}{\begin{figure}[tbp]
%\begin{center}
%\includegraphics[width=1.0\linewidth]{fig5.eps}
%\end{center}
%\caption{(a) Monolayer energy $E(N)/N$ and (b)
%corresponding surface energy for fully
%relaxed Cu(001) $1\times1$ slabs as
%a function of thickness. The red curve is a least-squares fit to the surface energy.}
%\label{fig5}
%\end{figure}}}%
%BeginExpansion
\begin{figure}[tbp]
\begin{center}
\includegraphics[width=1.0\linewidth]{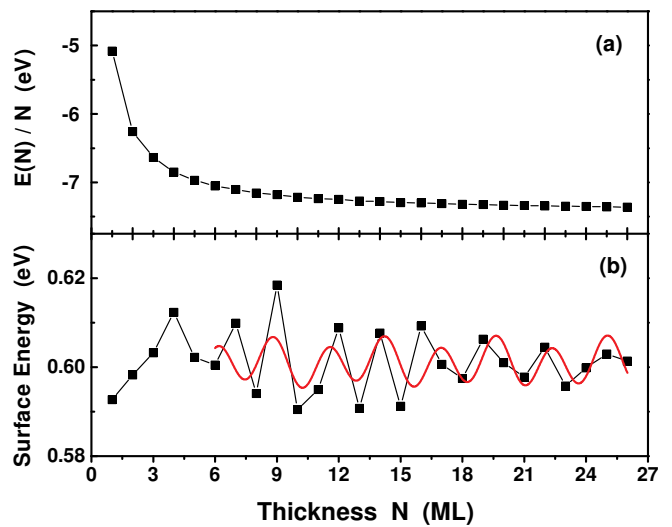}
\end{center}
\caption{(a) Monolayer energy $E(N)/N$ and (b)
corresponding surface energy for fully
relaxed Cu(001) $1\times1$ slabs as
a function of thickness. The red curve is a least-squares fit to the surface energy.}
\label{fig5}
\end{figure}%
%EndExpansion
An energetic quantity more suitably tailored to QSE is the surface energy
which is defined as one-half of the energy difference between the film and the
bulk with the same number of atoms, including the proper subtraction of a term
linear in $N$\cite{Boe2}. The thickness dependence of surface energy is
calculated and shown in Fig. 3(b) (curve with squares). It reveals that the
surface energy approximately follows a periodic oscillatory form, in accord
with the oscillation properties of film $D(E_{F})$. This can be simply
reasoned by the fact that the total band energy with respect to the Fermi
level is related to the DOS by the equation $E_{\text{band}}=\int^{E_{F}%
}(E-E_{F})D(E)dE$. Therefore, each time the total DOS crosses the Fermi level,
the band energy will also as a response have an oscillatory change, which
consequently results in periodic oscillations in surface energy. In the same
manner as discussed above, one can expect that the oscillations in the surface
energy consist of a superposition of oscillations with the periods
corresponding to the extremal points of the bulk Fermi surface. To illustrate
this, also plotted in Fig. 3(b) (red curve) is a least-squares fit to the
surface energy with the following expression%
\begin{equation}
E_{\text{surf}}(N)=\sum_{i=1}^{2}A_{i}\sin\left(  \frac{2\pi N}{\Lambda_{i}%
}+\phi_{i}\right)  , \tag{4}\label{E4}%
\end{equation}
where $N$ is the number of monolayers (ML), $\Lambda_{1}=5.6$ and $\Lambda
_{2}=2.7$ are the periodicities (in unit of atomic ML) corresponding to the
bell and neck extrema in bulk Cu Fermi surface, $A_{i}$ and $\phi_{i}$ are the
fitting parameters. It can be seen that the fitting curve well reproduces the
oscillating behavior in the surface energy. Thus the periods extracted from
the calculated Cu(001) surface energy are well consistent with the periods
extracted from the bulk Fermi surface, and again the stationary extrema in
Fermi surface play a key role in determining the oscillatory behavior in
surface energetics.%

%TCIMACRO{\TeXButton{TeX field}{\begin{figure}[tbp]
%\begin{center}
%\includegraphics[width=1.0\linewidth]{fig6.eps}
%\end{center}
%\caption
%{The second difference of the total energy as a function of thickness for freestanding Cu(001) thin films.}
%\label{sta}
%\end{figure}}}%
%BeginExpansion
\begin{figure}[tbp]
\begin{center}
\includegraphics[width=1.0\linewidth]{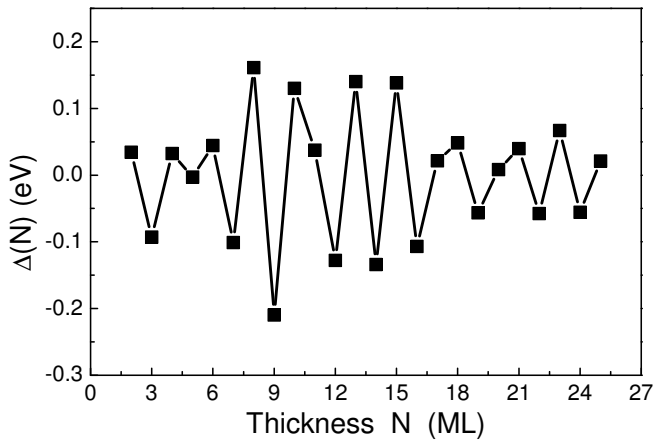}
\end{center}
\caption
{The second difference of the total energy as a function of thickness for freestanding Cu(001) thin films.}
\label{sta}
\end{figure}%
%EndExpansion
Some authors have used the energies of QW states at $\overline{\mathrm{\Gamma
}}$ to discuss the stability of the film. See Ref.\cite{Wei2003} for an
example. The key point employed there is that even though the confinement
takes place only in one of the three dimensions, the stability of the film
could also be affected due to the variation in the electronic
energy\cite{Zhang}. In the present case, since the above discussions have
clearly shown that the neck point $\mathrm{P}$ in the bulk Fermi surface plays
the same important role as the bell point (corresponding to $\overline
{\mathrm{\Gamma}}$ in surface BZ) does in determining the oscillatory
energetics properties of Cu(001) thin film, thus one can reasonably expect
that the stability of the film also exhibits a superposed oscillatory behavior
as a function of the film thickness, with the oscillation periods consisting
of long and short length. To illustrate this, following Ref.\cite{Wei2003}, we
use the second difference of the total energy as a measure of the relative
stability of an $N$-layer film with respect to the films of $N+1$ and $N-1$
layers, which is defined as
\begin{equation}
\Delta(N)=E(N+1)+E(N-1)-2E(N) \tag{5}\label{E5}%
\end{equation}
where $E(N)$ is the calculated total energy of the fully relaxed $N$-layer
film with the in-plane lattice constant fixed at the theoretical value. The
result is shown in Fig. 5. A peak in the figure indicates a high relative
stability for the film. It is not surprising that the stability is featured by
a superposition of long- and short-period oscillations as is evident from the
discussion above. Remarkably, one can see the $N=8$ film is mostly stable
while the $N=9$ film is particularly unstable.%

%TCIMACRO{\TeXButton{TeX field}{\begin{figure}[tbp]
%\begin{center}
%\includegraphics[width=1.0\linewidth]{fig7.eps}
%\end{center}
%\caption{(a) Work function for clean and Cs-adsorbed Cu(001)
%thin films as a function of thickness. (b) Planar-averaged electrostatic
%potential of clean (solid curve) and Cs-adsorbed (dotted curve) Cu(001) thin film (ten-layer Cu slab).}
%\label{fig6}
%\end{figure}}}%
%BeginExpansion
\begin{figure}[tbp]
\begin{center}
\includegraphics[width=1.0\linewidth]{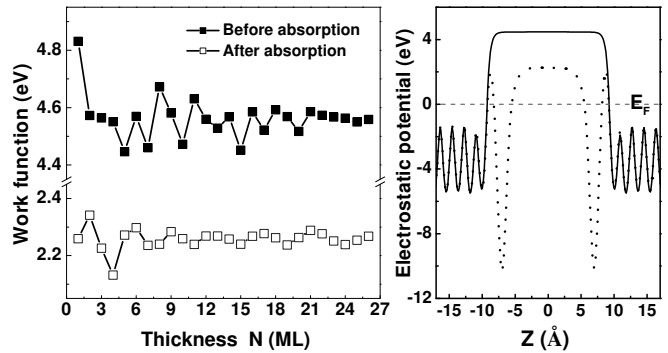}
\end{center}
\caption{(a) Work function for clean and Cs-adsorbed Cu(001)
thin films as a function of thickness. (b) Planar-averaged electrostatic
potential of clean (solid curve) and Cs-adsorbed (dotted curve) Cu(001) thin film (ten-layer Cu slab).}
\label{fig6}
\end{figure}%
%EndExpansion
In addition to the surface energy, we have also calculated the work function
$W$ of freestanding Cu(001) thin film. The work function, defined as the
minimum energy required to emit an electron from the surface to the vacuum, is
one fundamental physical quantity for surface reactivity. An elementary
picture of the work function involves a surface dipole layer that a valence
electron must overcome in order to escape. Since the work function, as with
many other properties, is also a function of the electronic density, thus the
changes in the electronic density by the crossings between the QW states and
Fermi level will influence the work function in an oscillatory way. In
addition to its relevance as an important element in our understanding of
surface science, a modified, or tunable, work function can be useful for
applications such as catalysis, because a slight change in the energy scale is
exponentially amplified for chemical reactions\cite{Tang2002}. Recent
\textit{in situ} experiments have measured atomic-layer-resolved work function
and shown clear QSE in Ag/Fe(100)\cite{Paggei2002} and
Pb/Si(111)\cite{Xue2007} systems. Here we have carried out first-principles
calculations of the work function of the clean Cu(001) surface. The result is
shown in Fig. 7(a) (filled squares) as a function of film thickness with
relaxed atomic geometry. One can see that the work function is featured by an
oscillatory behavior. As with the surface energy, the oscillations in the work
function of Cu(001) thin films consist of a superposition of long- and
short-length periods. \begin{table}[th]
\caption{Interlayer relaxations given in percent, $\Delta d_{i,i+1}$, of
Cu(001) as a function of the thickness of the film.}
%\label{specs}%
\begin{tabular}
[c]{ccccccc}\hline\hline
{$N$} & $\Delta d_{12}$ & $\Delta d_{23}$ & $\Delta d_{34}$ & $\Delta d_{45}$
& $\Delta d_{56}$ & $\Delta d_{67}$\\\hline
2 & -5.186 &  &  &  &  & \\
3 & -3.351 & -3.350 &  &  &  & \\
4 & -3.131 & +0.616 & -3.144 &  &  & \\
5 & -3.425 & +0.081 & -0.099 & -3.431 &  & \\
6 & -2.798 & +0.477 & -0.213 & +0.487 & -2.800 & \\
7 & -4.076 & +0.126 & +0.398 & +0.394 & -0.115 & -4.066\\
8 & -2.693 & +0.189 & -0.320 & -0.066 & -0.317 & +0.185\\
9 & -2.917 & +0.369 & +0.133 & +0.484 & +0.492 & +0.134\\
10 & -3.358 & +0.343 & -0.364 & +0.039 & -0.335 & -0.042\\
11 & -2.805 & +0.415 & +0.256 & +0.415 & -0.029 & -0.025\\
12 & -3.279 & +0.296 & -0.274 & +0.382 & -0.219 & -0.498\\
13 & -3.699 & +0.237 & -0.322 & -0.134 & -0.407 & -0.938\\
14 & -2.968 & +0.469 & +0.019 & +0.177 & -0.041 & +0.037\\
15 & -3.286 & +0.127 & -0.315 & -0.245 & -0.721 & -1.015\\\hline\hline
\end{tabular}
\end{table}%

%TCIMACRO{\TeXButton{TeX field}{\begin{figure}[tbp]
%\begin{center}
%\includegraphics[width=1.0\linewidth]{fig8.eps}
%\end{center}
%\caption{Interlayer relaxations of Cu(001) thin films as a function
%of the film thickness.} \label{fig7}
%\end{figure}}}%
%BeginExpansion
\begin{figure}[tbp]
\begin{center}
\includegraphics[width=1.0\linewidth]{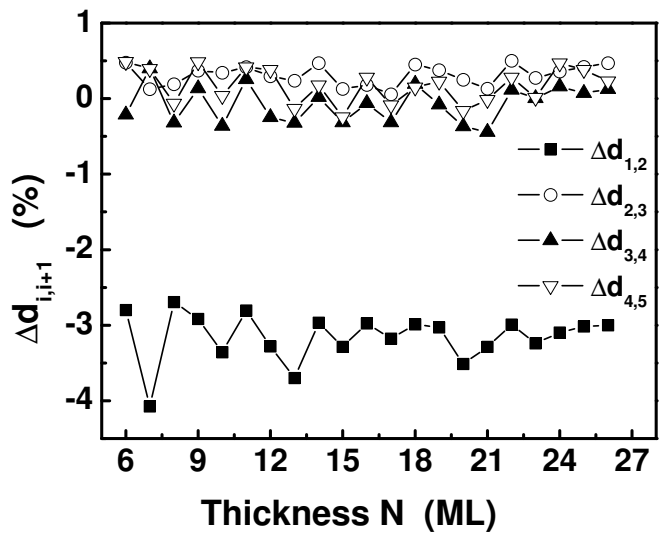}
\end{center}
\caption{Interlayer relaxations of Cu(001) thin films as a function
of the film thickness.} \label{fig7}
\end{figure}%
%EndExpansion
In addition to the above discussed film energetics, the relaxed atomic
structures of the Cu(001) thin film will also be influenced by the occurrence
of QW states at the Fermi level. Since the $1\times1$ supercells are employed
and $\mathbf{k}_{\parallel}$ is a good quantum number, thus only atomic
interlayer relaxation along [001] direction is allowed during our
calculations. Here the interlayer relaxation, $\Delta d_{i,i+1}$, is given in
percent with respect to the unrelaxed interlayer spacings, $d_{0}$, i.e.,
$\Delta d_{i,i+1}=100(d_{i,i+1}-d_{0})/d_{0}$. $d_{i,i+1}$ is the interlayer
distance between two adjacent layers parallel to the surface calculated by
total energy minimization. $d_{0}=a/2$ is the bulk interlayer distance along
[001] direction. Obviously, the signs $+$ and $-$ of $\Delta d_{i,i+1}$
indicate expansion and contraction of the interlayer spacings, respectively.
The interlayer relaxations of Cu(001) films as a function of the film
thickness is summarized in Table I. Furthermore, the interlayer relaxations
are also plotted in Fig. 8 as a function of $N$ for clear illustration. One
can see: (i) The two outmost layers relax significantly from the bulk value,
in agreement with the result from FLAPW calculation\cite{Silva02}. In the
whole range of layers that we considered, the topmost interlayer relaxation is
always inward ($\Delta d_{1,2}<0$), with $\Delta d_{1,2}$ starting from the
largest value of -5\% for a slab with only two monolayers, and approaches a
final value of -3\% with increasing the thickness of Cu(001) films. Whereas,
the second interlayer relaxation is always outward ($\Delta d_{2,3}>0$). Note
that the first interlayer separation on most metal surfaces is contracted,
Cu(001) is one of the typical examples. (ii) The interlayer spacings oscillate
as a function of the thickness of the film with the period again consisting of
long- and short-length scales. After 26 monolayers, which is the maximal
layers considered here, the oscillations are invisible which suggest that the
semi-infinite surface limit is now reached.

\section{Adsorption of cesium: QSE in surface cesiation}

In the above discussions we have extensively studied the QW states in Cu(001)
thin films and the corresponding QSE in various physical quantities such as
the surface energy, work function, and interlayer relaxation. To further
illustrate the physical properties influenced by finite size of the thin
films, in this section we focus our attention to the adsorption of Cs on
Cu(001) thin films. Note that our study in this section is not just for an
alternative verification of QSE in solid thin film. On the contrast, the
present study of surface adsorption as a function of the film thickness has
its own cause to be emphasized. It is well known that metal surfaces are a
prototype heterogeneous catalyst, and have been widely studied in terms of the
dissociative chemisorption of the reactant molecules. For a given solid,
except for surface irregularities such as steps and defects, its surface
reactivity is almost solely determined by the crystallographic orientation. At
reduced size and/or dimensionality, particularly when the characterized size
enters nanometer scale, however, the situation could be dramatically different
from that of the bulk. In fact, size-dependent surface chemical activities in
metal films with the thickness of nanometer scale have been observed in
previous experimental reports\cite{Danese2004,Aballe2004}. Remarkably, in a
very recent oxygen adsorption experiment on high-quality Pb(111) thin films,
Ma et al have clearly observed an oscillatory dependence of the chemical
reactivities on the film thickness\cite{Xue2007}, thus providing a most
convincing proof of the key role played by the QW states in changing surface
reactivity. It is expected that more experiments will address the issue of QSE
in surface chemical reactivities. From this aspect, the present detailed
theoretical analysis for the dependence of behavior of Cs adsorption upon the
Cu(001) film thickness is highly interesting and will be helpful for
experimental reference in near future.%

%TCIMACRO{\TeXButton{TeX field}{\begin{figure}[tbp]
%\begin{center}
%\includegraphics[width=0.50\linewidth]{fig9.eps}
%\end{center}
%\caption{The four different adsorption sites for Cs adatom on
%Cu(001) surface.} \label{fig8}
%\end{figure}}}%
%BeginExpansion
\begin{figure}[tbp]
\begin{center}
\includegraphics[width=0.50\linewidth]{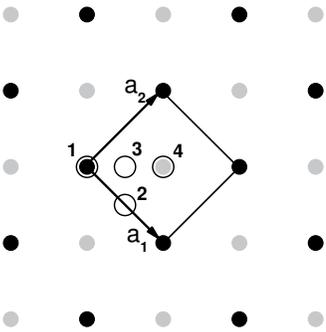}
\end{center}
\caption{The four different adsorption sites for Cs adatom on
Cu(001) surface.} \label{fig8}
\end{figure}%
%EndExpansion
Before we study the Cs adsorption properties as a function of the thickness of
Cu(001) thin films, we need to determine the energetically favorable
adsorption site. Since the preference of adsorption site is not sensitive to
the thickness of the substrate, thus to look for this preference of the
adsorption site, it is sufficient to give a study on the slabs with fixed
thickness of the Cu(001) substrate, which at present is chosen to be 5 ML. We
choose four most probable adsorption sites, which is enumerated in Fig. 9. The
binding energy is calculated using the following equation: Binding energy
[atomic Cs]$=-$($E[$Cs/Cu(001)$]-E[$Cu(001)$]-2E[$Cs$]$)/2 where
$E[$Cs/Cu(001)$]$ is the total energy of a slab which includes two Cs atoms
inside with a symmetric configuration, $E[$Cu(001)$]$ is total energy of the
slab without Cs atoms, and $E[$Cs$]$ is total energy of a free Cs atom which
is put in a 16 \AA $\times$16 \AA $\times$16 \AA supercell. The calculated
Cs/Cu(001) binding energies are 0.9189 eV, 0.9176 eV, 0.9174 eV, and 0.9327 eV
for site-1, site-2, site-3, and site-4 respectively. Thus site-4 is most
stable for adsorption. This is in accord with recent experiments that for
substrate surfaces with square or rectangular symmetry the alkali atoms occupy
adsorption sites which maximize the coordination number to the
substrate\cite{Berndt,Andersson,Demuth,Eggeling,Aminpirooz,Muschiol,Mizuno}.
Thus in the following the atomic Cs is always put on site-4 during the
simulation. This adsorption site is independent on the coverage $\Theta$ of Cs
atoms. In this paper, we only consider the case of $\Theta=0.5$.%

%TCIMACRO{\TeXButton{TeX field}{\begin{figure}[tbp]
%\begin{center}
%\includegraphics[width=1.0\linewidth]{fig10.eps}
%\end{center}
%\caption{(Color online) (a) The contour plot of charge density
%difference, $\rho($Cs$/$Cu(001)$)-\rho($Cu(001)$)-\rho($Cs$)$, in the (100)
%plane, for the cesiated Cu(001) surface. (b) (Left panel) band structure of
%cesiated Cu(001) thin film (5-layer Cu slab) with coverage $\Theta=0.5$;
%The right panel depicts the decomposition of local DOS for Cs
%adatom (dotted curves) and surface Cu atom (solid curves) into
%states with $s$ (cyan curves), $p$ (red curves), and $d$ (black curves)
%character for the Cs/Cu(001)system. The inset in (b) pedicts the
%band charge densities around the Fermi level (lower panel) and
%above the Fermi energy (upper panel) of typical value of 3.5 eV.} \label{fig9}
%\end{figure}}}%
%BeginExpansion
\begin{figure}[tbp]
\begin{center}
\includegraphics[width=1.0\linewidth]{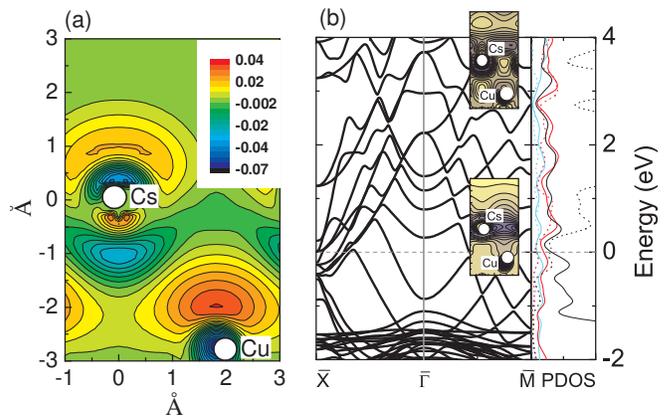}
\end{center}
\caption{(Color online) (a) The contour plot of charge density
difference, $\rho($Cs$/$Cu(001)$)-\rho($Cu(001)$)-\rho($Cs$)$, in the (100)
plane, for the cesiated Cu(001) surface. (b) (Left panel) band structure of
cesiated Cu(001) thin film (5-layer Cu slab) with coverage $\Theta=0.5$;
The right panel depicts the decomposition of local DOS for Cs
adatom (dotted curves) and surface Cu atom (solid curves) into
states with $s$ (cyan curves), $p$ (red curves), and $d$ (black curves)
character for the Cs/Cu(001)system. The inset in (b) pedicts the
band charge densities around the Fermi level (lower panel) and
above the Fermi energy (upper panel) of typical value of 3.5 eV.} \label{fig9}
\end{figure}%
%EndExpansion
To illustrate the surface bond formed between Cs adatom and Cu surface atoms,
we depict in Fig. 10(a) the contour map of the electron-density difference for
Cs/Cu(001). Here the plane depicted is normal to the surface and contains a Cs
and two Cu's. The density difference was obtained by subtracting the densities
of noninteracting component systems, $\rho($Cu(001)$)+\rho($Cs$)$, from the
density of the Cs/Cu(001) system, $\rho($Cs$/$Cu(001)$)$, while retaining the
positions of the component systems at the same location as Cs/Cu(001). The
solid and broken lines denote an increase and decrease, respectively, in
electron density upon Cs adsorption onto the surface. Covalent bonding is
evident from the accumulation of the charge between the Cs adsorbate and the
Cu(001) substrate. This charge is drawn principally from the atomic 6$s$ state
of Cs adlayer. On the vacuum side of the overlayer the charge is depleted.
Thus the Cs 6$s$ state causes a polarization toward the Cu(001) surface. As a
response, the work function will be decreased by this occurrence of surface
polarization. On the other side, a significant \textit{counter-polarization}
of the Cs 5p semicore charge is also quite evident from Fig. 10(a). This
effect opposes the reduction of the work function and the ultimate value of
work function represents a self-consistent compromise between 6$s$ and 5$p$
polarization. The net result of these multiple surface dipoles, as shown in
Fig. 7(b), is a lowering of the work function upon cesiation from 4.473 eV
[clean ten-layer Cu(001) slab] to 2.259 eV for the coverage $\Theta=0.5$,
corresponding to the relaxed height of the Cs atom above the Cu film of 2.988
\AA . Also due to this opposite orientation between 6$s$ and 5$p$
polarization, the work function of Cs/Cu(001) displays a non-monotonic
variation with increasing the coverage of Cs: At low coverage, the work
function decreases with increasing the coverage, going through a minimum
(about at $\Theta=0.5$) and increases a little bit from there on to the high
coverage value\cite{Arena1997}. The similar picture of cesiation process has
largely discussed in Cs/W(001) system\cite{Wimmer1983}. A well-know conclusion
in studying alkali-metal-atom chemisorption onto a metal surface is that
although the region around the alkali-metal adatom is electrically neutral and
no net charge transfer towards the metal surface is exhibited, the electrons
of the alkali-metal adatom undergo a strong mixing with the substrate electron
orbitals during the adsorption process. Such a process is local around the
adatom and accompanied by a screening process, which is responsible for work
function change. Furthermore, Figure 10(b) shows the band structure (left
panel) and the orbital-resolved local DOS (right panel) for the Cs adatom and
surface Cu atom, respectively. The complex Cs-induced charge rearrangement is
more obvious. The Cs adatom experiences during adsorption a repulsive
interaction of the valence electron with the induced image charge of the ionic
nuclear core. As a result, the valence electrons shift upward in energy and
hybridize with those of the substrate into bonding and antibonding
states\cite{Nordlander,Ishida}. The bonding states are mainly around the Fermi
level and are dominated by hybridization between adatom $sp$ and surface Cu
$sp$ orbitals, as shown in the lower panel in the inset in Fig. 10(b).
Whereas, the antibonding states shift up away from the Fermi level with a
typical value of 3.8 eV, and largely consists of hybridization between Cu $s$
and Cs $d$ orbitals, which can be seen from the upper panel in the inset in
Fig. 10(b). A thorough description of the bonding properties in Cs/Cu(001)
system is beyond our intention in this paper.%

%TCIMACRO{\TeXButton{TeX field}{\begin{figure}[tbp]
%\begin{center}
%\includegraphics[width=1.0\linewidth]{fig11.eps}
%\end{center}
%\caption{Calculated (a) binding energy of Cs adatom and (b) adsorbate
%height as a function thickness of Cu(001) films.} \label{fig10}
%\end{figure}}}%
%BeginExpansion
\begin{figure}[tbp]
\begin{center}
\includegraphics[width=1.0\linewidth]{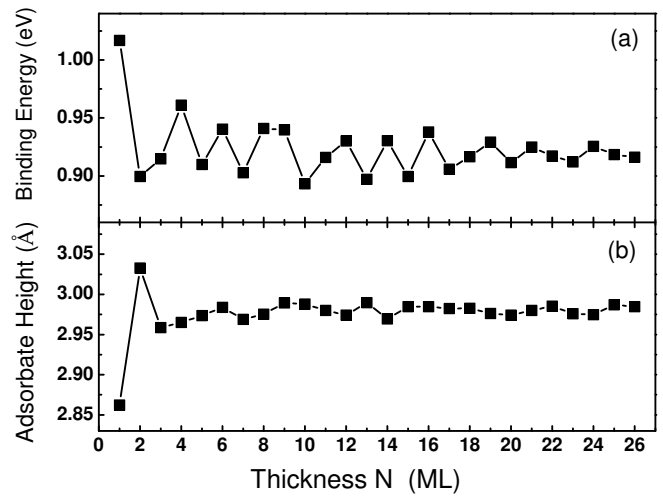}
\end{center}
\caption{Calculated (a) binding energy of Cs adatom and (b) adsorbate
height as a function thickness of Cu(001) films.} \label{fig10}
\end{figure}%
%EndExpansion
After finding the preferred atomic Cs adsorption site [site-4 in Fig. 9] and
getting familiar with the Cs-Cu(001) bonding properties, we turn now to our
central focus on the QSE in surface chemisorption and reactivity. For this
purpose, we have given a series of calculations for the binding energy (the
reverse of the adsorption energy) of the Cs adsorbate as a function of the
thickness of Cu(001) thin films. Here as mentioned above we only consider the
case of $\Theta=0.5$. The results are depicted in Fig. 11(a). One can see that
the binding energy of Cs onto Cu(001) thin films depends on the film thickness
in a damped oscillatory way. These oscillations are featured by a
superposition of long- and short-length periods, thus indicating a
well-defined QSE in the surface chemical reaction of Cu(001) thin film with
respect to Cs adsorption. In experiment this QSE of atomic adsorption can be
observed by investigating the dependence of Cs coverage on the monolayers of
Cu(001) thin films. We address here that the calculated data in Fig. 11 have
been carefully checked. This ensures that the oscillating behavior of the
adsorption energy has nothing to do with the convergence problem, which may be
encountered in the cluster calculations. The supercell\ approach retains 2D
periodic boundary conditions during calculation, and can avoid this problem.
Therefore, as with other quantities, the oscillations in adsorption energy is
physically caused by the periodic change in the DOS at $E_{F}$. In fact, from
what we have shown in Figs. 10, or from the simplest Anderson-Grimsley-Newns
adsoptin model\cite{Newns,Norskov}, one recalls that as a Cs atom approaches
to the Cu(001) surface (which can be approximated by a free-electron-like
metal), the valence $s$-$p$ states of Cs adatom are broaden into resonant
states due to as a result of interaction with the metal bands. Surface
electrostatic potential will shift this hybridization down in energy to below
$E_{F}$ for sustaining the whole neutralization. A lower DOS at $E_{F}$
implies that the Cu(001) film has fewer electronic states to respond, whereas
a higher $D(E_{F})$ means the film has more electronic states to respond, to
the presence of the Cs adsorbate. Therefore, higher $D(E_{F})$ implies higher
probability in the above hybridization process, causing the $sp$ resonance to
move to lower energies and to be occupied. This leads to a higher surface
reactivity and higher adsorption energy. This physical picture can be
formulated by expressing the adsorption energy as
\begin{align}
E_{\text{ad}}  &  =E_{\text{band}}(\text{un-adsorbed})-E_{\text{band}%
}(\text{adsorbed})\tag{6}\label{E6}\\
&  \approx\int_{-\infty}^{E_{F}}E\Delta DdE,\nonumber
\end{align}
where $E_{\text{band}}($un-adsorbed$)$ is the band energy for a combined
Cs/Cu(001) system with Cs adatom and Cu(001) film largely separated such that
no chemsorption occurs, $E_{\text{band}}($adsorbed$)$ is the band energy for
the adsorbed Cs/Cu(001) system, $\Delta D=$DOS$($un-adsorbed$)-$
DOS$($adsorbed$)$ is the difference in the electronic DOS between the
un-adsorbed and adsorbed Cs/Cu(001) systems. As the film thickness changes, QW
state energy levels shift. Whenever a QW state crosses the Fermi level from
above, it adds energy to the $E_{\text{band}}($un-adsorbed$)$ and adds
electronic density to un-adsorbed $D(E_{F})$. Whereas $E_{\text{band}}%
($adsorbed$)$ decreases upon the crossing between the QW states and the Fermi
level due to the above mentioned cause that the Cs-Cu orbital hybridization
shifts more down in energy by the increase of $D(E_{F})$. Thus the net
consequence is an anti-phase oscillation mode of the adsorption energy with
respect the oscillation mode of $D(E_{F})$ as a function of the film thickness.

The Cs adsorbate height is plotted in Fig. 7(b) as a function of the film
thickness, which also shows the periodic oscillations indicative of QSE.
Compared to the features in thickness dependence of binding energy, one can
notice that oscillations in the adsorbate height are much weak and the periods
are somewhat not easily resolved. The cause may be due to the fully interlayer
relaxations as a whole structure during calculations. If the substrate is not
allowed to relax during the adsorption process, then it can be expected that
there is a consistent correspondence in the oscillations between the binding
energy and the height of the adsorbate.

Moreover, we have also calculated the work function of Cs/Cu(001) as a
function of the film thickness. The results are shown in Fig. 7 (open
squares). Compared to the case of clean Cu(001) films, the work function is
decreased by a typical value of 2.2 eV [see Fig. 7(b) for 10-layer case].
Also, the adsorbed work function oscillates with increase of the film
thickness. However, one can see that the oscillating mode is different from
that without adsorption, and also the oscillating amplitude becomes much weak.
This implies that the QW states are different for clean and cesiated Cu(001)
thin films. As a result, there no longer exists a simple one-to-one
correspondence between the adsorbed QW states and the bulk Cu Fermi surface.
More detailed work will be done for this issue.

\section{Conclusion}

In summary, the clean and cesiated Cu(001) thin films have been extensively
studied by density-functional theory pseudopotential plane-wave calculations.
The dependence of electronic structure, surface energetics, and interlayer
relaxation upon the thickness of the film has been fully investigated, clearly
showing the metallic QSEs of the film. These QSEs have been shown to be
closely related with the occurrence of QW states at the Fermi level. Unlike
some other simple $sp$ metals such as Al, Mg, and Pb, the Fermi surface of
bulk Cu is characterized by multiple stationary extrema such as the belly and
the neck points. As a consequence, the Cu(001) QW states also display the
abundant properties at these surface-projected $k$-extrema. For example, we
have shown the different film-thickness oscillation modes by calculating the
energies of QW states in correspondence with these stationary extrema. Due to
the interference between these two kinds QW states, the energetics and the
corresponding stability (Fig. 6) of the Cu(001) thin films have been shown in
a consistent way to display a quantum beating behavior as a function of the
film thickness, with the oscillation periods consisting of long and short
length. In addition, the calculated energy gap between the highest occupied QW
state and lowest unoccupied QW state, which can be directly measured via STS
technique, have been shown to display different slopes and kinks at these
extrema in bulk BZ.

We have also extensively discussed the other highly interested issue, i.e.,
the oscillatory QSE in surface chemical reactivities, via studying the Cs
adsorption onto the Cu(001) thin films. Through systematic calculations, we
have shown that the Cu(001) surface chemical catalysis for Cs adsorption can
be uniquely modulated and controlled by the occurrence of QW states at the
Fermi level. The Cs adsorption energy and the work function after cesiation
are featured by quantum oscillations as a function of the film thickness. The
amplitudes in these oscillations reach a typical value of 40 meV. From the
application point of view, this result is quite intriguing because 1ML change
in the film thickness can cause an effect equivalent to a temperature change
as high as 400 degrees. These calculated results thus may be used as a guide
to tailor catalysis, chemical reactions and other surface processes in
nanostructured materials.

\begin{acknowledgments}
This work was partially supported by the CNSF under Grant No. 10544004 and 10604010.
\end{acknowledgments}

\end{document}